\documentclass[modern]{aastex631}
\usepackage[utf8]{inputenc}
\usepackage{amsmath}
\usepackage{amssymb}
\usepackage{graphicx}
\usepackage{xcolor}
\usepackage{natbib}
\usepackage{hyperref}
\usepackage{comment}

\begin{document}

\title{\textbf{Bound Domains}}

%\correspondingauthor{G. M. Voit}

\author[0000-0002-3514-0383]{G. Mark Voit}
\affiliation{Michigan State University,
Department of Physics and Astronomy,
East Lansing, MI 48824, USA}

\begin{abstract}
% \begin{center} 
%    \quad \marknotes{Abstract goes here}
% \end{center}
How much energy is required to unbind baryons from the cosmological structures that originally bind them? This tutorial article explains why trying to answer this question using just a halo model can be misleading. Instead, it recommends parsing the universe into ``bound domains,'' which are the gravitationally bound structures that ultimately become widely separated islands as the universe evolves. It explains why a bound domain's potential well was about as deep $\sim 1$~Gyr after the Big Bang as it is now, and it outlines how future research might take advantage of a bound-domain approach to make progress on some open questions about the baryon distributions in and around galaxy groups and clusters.
\end{abstract}

\section{Introduction}

As an expanding $\Lambda$CDM universe with constant dark energy density proceeds into the far future, its cosmological halos collect into gravitationally bound islands that zoom away from each other with increasing speed \citep[e.g.,][]{Chiueh_2002PhRvD..65l3518C,NagamineLoeb_2003NewA....8..439N,Busha_2003ApJ...596..713B,Busha_2007ApJ...665....1B,Dunner_2006MNRAS.366..803D,Dunner_2007MNRAS.376.1577D}. As far as I know, those islands do not have a commonly used name, so I will call them \textit{bound domains.} 

I became interested in bound domains while thinking about how early feedback outbursts from supermassive black holes are connected to the atmospheres of present-day galaxy groups and clusters. Observational baryon inventories currently indicate that baryonic mass fractions within groups are about half the cosmic mean, at most, while baryon fractions in clusters appear nearly equal to the cosmic mean \citep{Eckert_2021Univ....7..142E}. And yet, clusters are thought to assemble through mergers of groups. The potential well of all the matter that will eventually form a galaxy cluster must therefore be able to retain most of the baryons that black-hole feedback may have pushed out of its progenitor groups earlier in time. But how?

In pursuit of more insight into how clusters retain their baryons and to assess how far baryons might go when they are pushed out of groups, this tutorial article outlines the properties of the bound domains in which galaxy groups and clusters grow and suggests some future research directions. Its purpose is to draw attention to how bound domains shape the universe's large-scale baryon distribution and to advocate for further investigation of their role using cosmological numerical simulations.

\section{Bound Domain Dynamics}
\label{sec:Dynamics}

To illustrate how a bound domain evolves, we will start with an idealized model, consisting of spherically symmetric and concentric mass shells that each follow the equation of motion
\begin{equation}
    \ddot{R} = - \frac {G M} {R^2} + H_0^2 \Omega_\Lambda R
    \; \; .
\end{equation}
The first term on the right is the Newtonian gravitational force exerted by the mass $M$ within a shell of radius $R$. The second one accounts for dark energy with a constant energy density that is $\Omega_\Lambda c^2$ times the current critical density $3 H_0^2 / 8 \pi G$. A non-relativistic approach suffices as long as $\dot{R} \ll c$ and spatial curvature is negligible.

Integrating $d \dot{R}^2 / dt$ over time using this non-relativistic expression for $\ddot{R}$ gives the standard result
\begin{equation}
    \frac {\dot{R}^2} {2} 
        = \frac {G M} {R} + \frac {H_0^2 \Omega_\Lambda} {2} R^2
            + \varepsilon_M
            \; \; ,
            \label{eq:EqOfMotion}
\end{equation}
in which $\varepsilon_M$ is a constant of integration related to the specific energy of matter in the shell encompassing mass $M$. Without dark energy, $\varepsilon_M$ is simply the shell's specific binding energy: The shell is gravitationally bound for $\varepsilon_M < 0$ and is unbound for $\varepsilon_M > 0$. 

However, a marginally bound shell in a gravitating system that includes dark energy has $\varepsilon_M < 0$. Both $\ddot{R}$ and $\dot{R}$ asymptotically approach zero as the shell approaches a finite radius. That radius is the limiting radius of a bound domain containing mass $M_{\rm bd}$. Setting $\ddot{R} = 0$ gives the limiting radius as a function of $M_{\rm bd}$:
\begin{equation}
    R_\infty = \left( \frac {G M_{\rm bd}} {H_0^2 \Omega_\Lambda} \right)^{1/3}
    \; \; .
\end{equation}
A marginally bound shell also needs to have $\varepsilon_M$ equal to 
\begin{equation}
    \varepsilon_\infty
        \: \equiv \: - \frac {3} {2} \frac {G M_{\rm bd}} {R_\infty}
        \: = \: - \frac {3} {2} \left( G M_{\rm bd} H_0 \right)^{2/3} \Omega_\Lambda^{1/3}
\end{equation}
so that $\dot{R}$ goes to zero at $R_\infty$. The equation of motion for the radius $R_{\rm bd}$ of the bound domain's outer edge is therefore
\begin{equation}
    \dot{R}_{\rm bd} 
      \: = \: \left[ 2 G M_{\rm bd} \left( \frac {1} {R_{\rm bd}} - \frac {1} {R_\infty}  \right) 
            + H_0^2 \Omega_\Lambda \left( R_{\rm bd}^2 - R_\infty^2 \right) \right]^{1/2}
            \; \; .
\end{equation}
Anything within $R_{\rm bd}$ that gains enough energy to move outside of $R_\infty$ is no longer gravitationally bound to the bound domain.

\section{Cosmological Potential Well}
\label{sec:Well}

Conveniently, this spherical bound domain model provides an unambiguous zero point for the gravitational potential. For example, the gravitational potential well of the spherical system we have been considering is 
\begin{equation}
    \varphi (R) \: = \: \int \frac {G M(R)} {R^2} dR 
                        \: - \frac {H_0^2 \Omega_\Lambda } {2} R^2 
                        \: + \: \varphi_0
                        \; \; ,
\end{equation}
where $\varphi_0$ is a constant of integration that provides flexibility to choose the radius at which $\varphi = 0$. However, the divergent dark-energy term makes it impossible to set this expression for $\varphi(R)$ equal to zero at infinity. Also, the limits of integration in the Newtonian term are unspecified, and it diverges if $M(R)$ increases more rapidly than $M \propto R$.

However, if we are mainly interested in whether a particular particle inside a bound domain is gravitationally bound to it, then we can isolate the bound domain's potential well from the rest of the universe. To illustrate how that works, we will treat the bound domain as a spherical system of discrete mass shells, in which shell $i$ has a radius $R_i$, encompasses a mass $M_i$, and obeys the equation of motion 
\begin{equation}
        \ddot{R_i} = - \frac {G M_i} {R_i^2} + H_0^2 \Omega_\Lambda R_i
                    \; \; .
\end{equation}
The Newtonian contribution of a particular mass shell to the overall potential well is then
\begin{equation}
    \varphi_i (R) 
        \: = \: \frac {G (M_i - M_{i-1})} {R_\infty}
                    \left[ 1 - \frac {R_\infty} {\max(R_i,R)} \right]
                    \; \; .
                    \label{eq:phi_i}
\end{equation}
Notice that the constant term places the zero point of this potential well at $R_\infty$. Likewise, we can represent the contribution of dark energy using
\begin{equation}
    \varphi_\Lambda (R) = \frac {H_0^2 \Omega_\Lambda } {2} \left( R_\infty^2 - R^2 \right) 
\end{equation}
which also places the zero point at $R_\infty$. With those choices, the bound domain's potential well becomes
\begin{equation}
    \varphi_{\rm bd} (R) = \varphi_\Lambda (R) + \sum_i  \varphi_i(R)
    \; \; .
\end{equation}
Lifting a particular particle in this potential well from radius $R$ to the radius $R_\infty$ at which it cannot remain bound therefore requires a specific energy equivalent to $- \varphi_{\rm bd}(R)$.

\section{Potential Well Evolution}
\label{sec:Evolution}

A bound domain's potential well starts out rather simple, when isolated from the rest of the universe, because all the matter belonging to the bound domain can be treated as a point mass as long as $R \gg R_{\rm bd}$. In that case, the bound domain's overall potential well is
\begin{equation}
    \varphi_{\rm bd,0}(R) 
        \: = \: - \frac {G M_{\rm bd}} {R}
                    \left( 1 - \frac {R} {R_\infty} \right)
                + \frac {H_0^2 \Omega_\Lambda} {2}
                    \left( R_\infty^2 - R^2 \right)
                \label{eq:phi_bd,0}
            \; \; .
\end{equation}
This potential well can be very deep early in time, when all the matter is concentrated within a region much smaller than $R_\infty$, but the bound domain's mass shells are destined for significant expansion because they are traveling outward with a specific kinetic energy only slightly less than $- \varphi_{\rm bd} (R_{\rm bd})$.

Each of the bound domain's mass shells eventually reaches a turnaround radius $R_{{\rm ta},i}$ at a turnaround time $t_{{\rm ta},i}$, after which the shell falls back inward. After shell $i$ expands beyond a fixed physical radius $R$, its contribution to the gravitational potential at $R$ is no longer like that of a point mass. Instead, its contribution to the potential at $R$ declines as $R_i$ increases, as expressed by equation (\ref{eq:phi_i}), because it no longer hinders a particle's ascent from $R$ to $R_i$. Expansion of the shell system as time progresses therefore makes it easier for energetic particles to escape a bound domain's innermost regions. However, the potential well of a particular shell deepens again after it reaches $R_{{\rm ta,}i}$ and reverses its trajectory.

Formally, an idealized spherical mass shell returns to the origin at a collapse time equal to $2 t_{{\rm ta},i}$. More realistically, though, the infalling matter becomes incorporated into a cosmological halo at that time. The rest of this article will therefore treat matter that has collapsed into the central halo differently from matter in mass shells that have not yet collapsed. 

Suppose that $M_{\rm halo} (t)$ describes the mass accretion history of a cosmological halo at a bound domain's center and that mass shell $j$ is the one that has most recently collapsed. The potential well of the bound domain at time $t$ is then
\begin{equation}
    \varphi_{\rm bd} (R,t)
        = \varphi_\Lambda (R) 
            + \varphi_{\rm halo} (R,t)
            + \varphi_{\rm shells} (R,t)
            \; \; ,
\end{equation}
where $\varphi_{\rm halo} (R,t)$ represents the potential well of a cosmological halo of mass $M_{\rm halo} = M_j$ at time $t$ and the collective gravitational potential of the mass shells that have not yet collapsed is
\begin{equation}
    \varphi_{\rm shells} (R,t)
        = \sum_{i > j(t)} \varphi_i (R,t)
            \; \; .
\end{equation}
The relative contributions of the halo and shells to the overall gravitational potential shift with time: $\varphi_{\rm halo}$ becomes deeper as $M_{\rm halo}$ grows, and $\varphi_{\rm shells}$ becomes shallower as $M_{\rm shells} = M_{\rm bd} - M_{\rm halo}$ declines.
Also, the halo's potential well depends somewhat on the halo's mass accretion history, which determines how centrally concentrated the halo's mass distribution is. 

\section{A Group-Scale Example}
\label{sec:Example}

Presenting an example will help make the discussion more concrete. First, we need a mass-accretion history for the bound domain's central halo as well as a total mass for the bound domain. We will express the halo's mass-accretion history using the formula
\begin{equation}
    M_{\rm halo} (t) = M_0 ( 1 + z )^\alpha \exp (\beta z)
\end{equation}
in which $z$ represents the cosmological redshift corresponding to time $t$. Both $\alpha$ and $\beta$ are model parameters that determine the rate of halo mass growth, and $M_0$ is the halo's mass at $z=0$. Following \citet{Correa_I_2015MNRAS.450.1514C,Correa_II_2015MNRAS.450.1521C}, we will adopt $\alpha = 0.24$ and $\beta = -0.75$. Extending a mass-accretion history with these values of $\alpha$ and $\beta$ into the future shows that $M_{\rm halo}$ peaks at  $1.27 M_0$, and so we will take $M_{\rm bd} = 1.27 M_0$ to be the mass of the bound domain in which this particular evolving halo resides. 

\begin{figure}[t]
    \centering
    \includegraphics[width=0.9\linewidth]{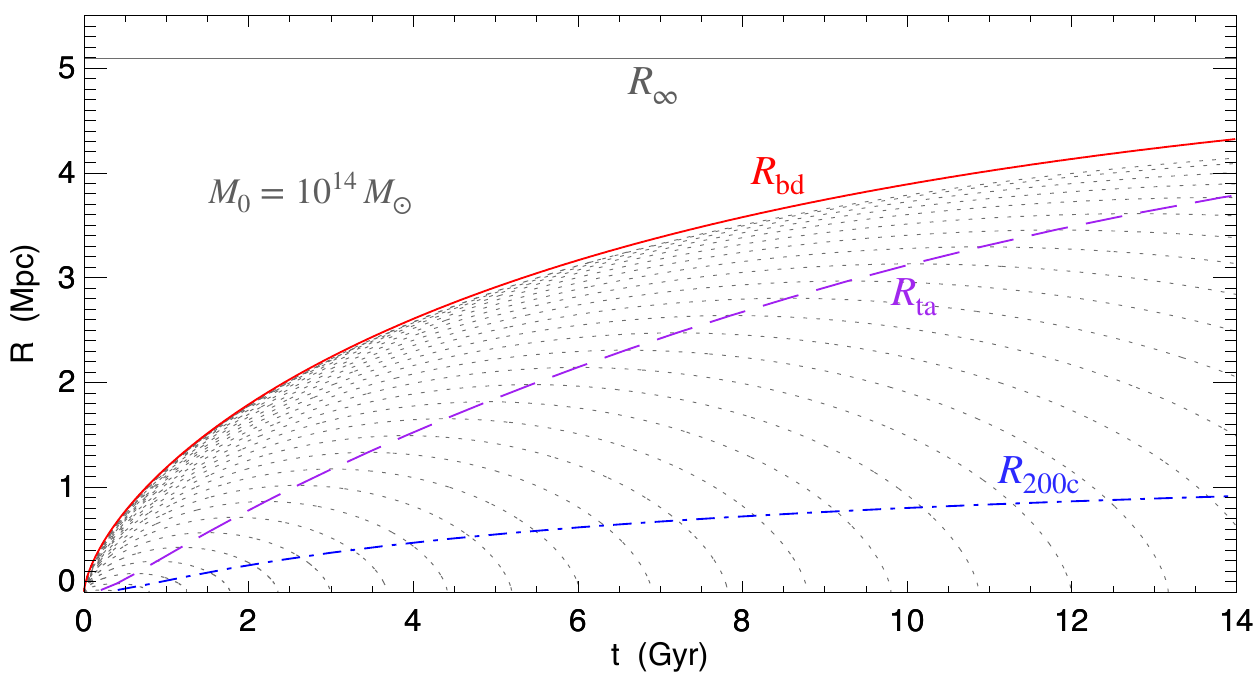}
    \caption{Cosmological evolution of a spherical bound domain centered on a group-scale halo with present-day mass $M_0 = 10^{14} \, M_\odot$. Its outer radius $R_{\rm bd}$ (solid red line) asymptotically approaches $R_\infty$ (solid grey line). Dotted grey lines show the trajectories of mass shells within $R_{\rm bd}$. Each one eventually reaches a turnaround radius $R_{\rm ta}$ (dashed purple line) and falls back into a central halo of radius $R_{\rm 200c}$ (dot-dashed blue line).}
    \label{fig:bd_evol_labeled}
\end{figure}

Figure \ref{fig:bd_evol_labeled} shows the evolution of such a bound domain around a halo with a present-day mass of $M_0 = 10^{14} \, M_\odot$, representing the high end of the mass range for galaxy groups. A solid red line shows how the outer radius $R_{\rm bd}$ of the halo's bound domain approaches its asymptotic radius $R_\infty = 5.1 \, {\rm Mpc}$, given by the black line. Dotted grey lines illustrate the trajectories of mass shells within the halo's bound domain. A dashed purple line shows how the turnaround radius $R_{\rm ta}$ of the mass shells evolves. A dot-dashed blue line shows how the central halo's virial radius grows as mass accretes onto it. In this illustration, the virial radius is equal to $R_{\rm 200c}$, defined to contain a mean matter density 200 times the universe's critical density in a flat $\Lambda$CDM universe with $\Omega_\Lambda = 0.7$.

To compute the bound domain's overall potential well, we need an expression for $\varphi_{\rm halo} (R)$. Here, we will approximate the halo's potential well using a Hernquist halo model \citep{Hernquist_1990ApJ...356..359H}:
\begin{equation}
    \varphi_{\rm halo} (R) 
        = \frac {G M_{\rm halo}} {R_\infty + R_{\rm H}}
          - \frac {G M_{\rm halo}} {R + R_{\rm H}}
    \; \; .
\end{equation}
Again, we include a constant term that makes $\varphi_{\rm halo} (R_\infty) = 0$. The circular velocity profile of this halo model peaks at the radius $R_{\rm H}$, and we set $R_{\rm H}$ equal to $0.5 R_{\rm 200c}$. The Hernquist halo's potential is then similar to that of an NFW halo \citep{NFW_1997ApJ...490..493N} with a scale radius $R_{\rm 200c} / 4.3$, but the NFW halo's potential does not converge toward a point-mass potential at $R \gg R_{\rm 200c}$. That is why a Hernquist halo model is superior for this illustration.

\begin{figure}[!t]
    \centering
    \includegraphics[width=0.9\linewidth]{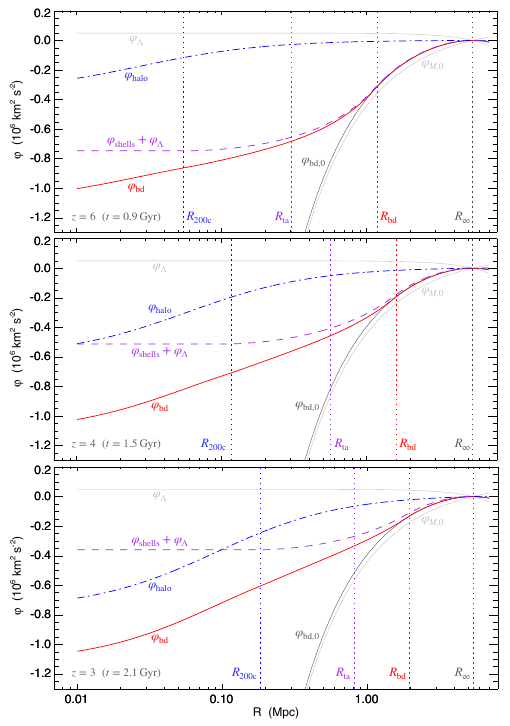}
    \caption{Evolution from $z=6$ to $z=3$ of the gravitational potential of the bound domain in Figure \ref{fig:bd_evol_labeled}. In each panel, a solid black line shows the initial potential well $\varphi_{\rm bd,0}$ consisting of the sum of the dark-energy potential $\varphi_\Lambda$ (grey line) and the potential well $\varphi_{M,0}$ of a point-like mass $M$ (another grey line). A dot-dashed blue line shows the potential well $\varphi_{\rm halo}$ of the bound domain's central halo, and a dashed purple line shows the sum of $\varphi_\Lambda$ and the collective potential well $\varphi_{\rm shells}$ of all the mass shells that have not yet fallen into the central halo. The bound domain's potential well $\varphi_{\rm bd}$ (solid red line) is the sum of $\varphi_{\rm halo}$, $\varphi_{\rm shells}$, and $\varphi_\Lambda$. Dotted vertical lines show $R_{\rm 200c}$, $R_{\rm ta}$, $R_{\rm bd}$, and $R_\infty$, as labeled.}
    \label{fig:bd_potential_hi-z}
\end{figure}

\begin{figure}[!t]
    \centering
    \includegraphics[width=0.9\linewidth]{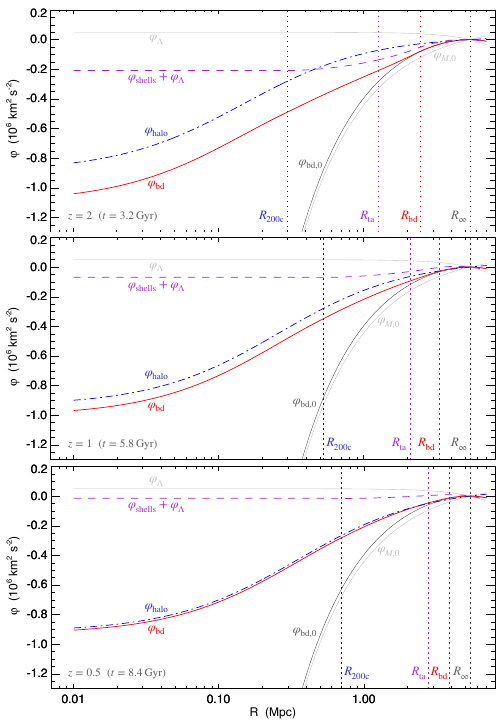}
    \caption{Evolution from $z=2$ to $z=0.5$ of the gravitational potential of the bound domain in Figure \ref{fig:bd_evol_labeled}. In each panel, a solid black line shows the initial potential well $\varphi_{\rm bd,0}$ consisting of the sum of the dark-energy potential $\varphi_\Lambda$ (grey line) and the point-mass potential well labeled $\varphi_{M,0}$ (another grey line). A dot-dashed blue line shows the potential well $\varphi_{\rm halo}$ of the bound domain's central halo, and a dashed purple line shows the sum of $\varphi_\Lambda$ and the collective potential well $\varphi_{\rm shells}$ of all the mass shells that have not yet fallen into the central halo. The bound domain's potential well $\varphi_{\rm bd}$ (solid red line) is the sum of $\varphi_{\rm halo}$, $\varphi_{\rm shells}$, and $\varphi_\Lambda$. Dotted vertical lines show $R_{\rm 200c}$, $R_{\rm ta}$, $R_{\rm bd}$, and $R_\infty$, as labeled.}
    \label{fig:bd_potential_lo-z}
\end{figure}

Figures \ref{fig:bd_potential_hi-z} and \ref{fig:bd_potential_lo-z} show how the resulting potential well of the bound domain in Figure \ref{fig:bd_evol_labeled} evolves. The main thing to notice is that the gravitational potential difference between $R \approx 10 \, {\rm kpc}$ and the potential maximum at $R_\infty$ remains nearly constant from $z \approx 6$ to the present day. Before $z \approx 4$, mass shells that have not yet fallen into the halo are responsible for most of the gravitational potential difference that confines matter within the bound domain. The potential well of the uncollapsed shells remains significant until $z \lesssim 2$, but the majority of the bound domain's mass has fallen into the central halo by $z \approx 1$, and so $\varphi_{\rm bd} \approx \varphi_{\rm halo}$ after that time.

These features of a bound domain's potential well depth have important implications for the baryons that feedback pushes out of a halo early in time. For example, this particular central halo has a mass $M_{\rm halo} \approx 2 \times 10^{12} \, M_\odot$ at $z \approx 6$, and escape velocity from its potential well at that moment is 
\begin{equation}
    v_{\rm esc} 
        \sim \sqrt{2 | \varphi_{\rm halo} |} 
        \sim 700 \, {\rm km \, s^{-1}}
        \; \; .
\end{equation}
Unbinding of the halo's baryons from the halo would therefore seem to require an energy $f_{\rm b} M_{\rm halo} |\varphi_{\rm halo}| \sim 10^{60} \, {\rm erg}$. However, unbinding of the halo's baryons from the \textit{entire} bound domain requires about twice the escape velocity and four times the energy. 

Furthermore, the baryons outside of $R_{\rm 200c}$ pose additional resistance. As ejected halo baryons move beyond $R_{\rm 200c}$, they encounter a headwind associated with infalling baryons between $R_{\rm 200c}$ and $R_{\rm ta}$. And beyond $R_{\rm ta}$ are even more baryons belonging to decelerating mass shells extending out to $R_{\rm bd}$. At this particular moment $(z \approx 6)$, the baryonic mass between $R_{\rm 200c}$ and $R_{\rm ta}$ is $\sim 5$ times the halo's baryonic mass. That factor rises to $\sim 50$ when the baryons out to $R_{\rm bd}$ are included. Section \ref{sec:BaryonBinding} will show that unbinding \textit{all} of those baryons requires $\sim 10^{62} \, {\rm erg}$ of energy input.

It therefore seems likely that halo baryons pushed beyond $R_{\rm 200c}$ by early feedback events will accumulate in the region between $R_{\rm 200c}$ and $R_{\rm ta}$. In principle, baryons that stop moving outward before reaching $R_{\rm ta}$ can later reaccrete onto the halo along with the infalling shells of cospatial collisionless matter, as long as those baryons follow nearly ballistic trajectories. But baryons are collisional, and continual injection of feedback energy into the bound domain might prevent baryons pushed beyond $R_{\rm 200c}$ from falling back inward. Either way, though, it is incorrect to consider baryons that early feedback has lifted out of a halo gravitationally unbound, and their lingering presence between $R_{\rm 200c}$ and $R_{\rm ta}$ may well influence what goes on inside of $R_{\rm 200c}$.

The bottom line is this: What happens to baryons between $R_{\rm 200c}$ and $R_{\rm ta}$ around galaxy groups is currently unclear. Clarifying what happens to them will require attention to the \textit{entire} gravitational potential well of the group's bound domain and how it evolves, not just the potential well of its central halo.

\section{Perturbation versus Background}
\label{sec:PertVsBack}

The near invariance of a bound domain's potential well depth while its central halo's mass grows by a factor of $\sim 50$ may at first seem paradoxical, since the amplitude of the matter-density perturbation in that region of space is clearly growing by orders of magnitude. To resolve the paradox, we will take a closer look at how both the universe's background matter density and the growing perturbation contribute to the bound domain's overall potential well.

\begin{figure}[!t]
    \centering
    \includegraphics[width=0.8\linewidth]{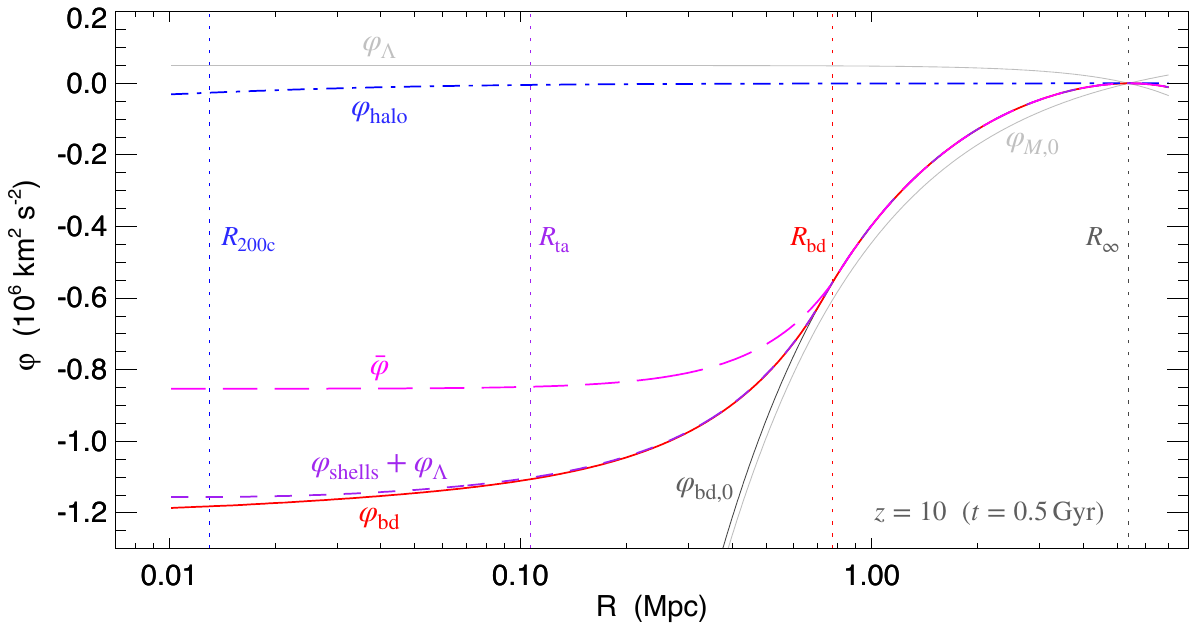}
    \caption{Comparison of the gravitational potential of the bound domain in Figure \ref{fig:bd_evol_labeled} with the background potential at $z=10$. A long-dashed magenta line shows the potential well $\bar{\varphi}(R)$ from equation (\ref{eq:phi_uniform}), representing an expanding uniform sphere with the background density. All other lines are identical to their counterparts in Figures \ref{fig:bd_potential_hi-z} and \ref{fig:bd_potential_lo-z}. At this early time, the bound domain's potential well depth is not much greater than the potential well of an unperturbed region of the universe with an identical mass.}
    \label{fig:bd_potential_phibar}
\end{figure}

First, the background: Consider a uniform sphere with the same mass as the bound domain and a density that is always equal to the universe's mean matter density $\bar{\rho} = 3 H_0^2 \Omega_M (1+z)^3 / 8 \pi G$. That sphere extends out to the radius
\begin{equation}
    \bar{R} \equiv \left( \frac {3 M_{\rm bd}} {4 \pi \bar{\rho}} \right)^{1/3}
            = \left( \frac {G M_{\rm bd}} {2 H_0^2 \Omega_M} \right)^{1/3} (1 + z)^{-1}
    \; \; . 
\end{equation}
Outside of $\bar{R}$, the sphere's potential well is just like the point-mass potential $\varphi_{\rm bd,0} (R)$ in equation (\ref{eq:phi_bd,0}). At smaller radii ($R < \bar{R}$), its potential well is 
\begin{equation}
    \bar{\varphi} (R) 
        = \left[ \frac {\Omega_M} {2} (1 + z)^3 - \Omega_\Lambda \right]
           \frac {H_0^2 (R^2 - \bar{R}^2)} {2} + \varphi_{\rm bd,0} (\bar{R})
    \label{eq:phi_uniform}
    \; \; .
\end{equation}
Figure \ref{fig:bd_potential_phibar} shows what the uniform sphere's potential well is like at $z = 10$. Its depth at that early time is nearly as great as the bound domain's potential well depth but is decaying in proportion to $1+z$.

Next, the perturbation: Inside of $\bar{R}$, the gravitational potential difference between $R$ and $\bar{R}$ depends on how perturbation growth alters the radii of the bound domain's mass shells. As time progresses, the radius $R_i$ of the shell encompassing mass $M_i$ lags behind the radius $\bar{R}_i$ the shell would have in an unperturbed universe. The gravitational potential of shell $i$ is different from the background case only at $R < \bar{R}_i$, and the difference becomes significant when $R_i$ becomes significantly less than $\bar{R}_i$. That happens as shell $i$ approaches its turnaround radius. After the shell turns around and falls back inward, its potential well becomes more and more like a point-mass potential. However, virialization prevents the infalling matter from returning to the origin, and so the bound domain's potential well does not become singular there. 

A bound domain's overall potential well depth is therefore determined primarily by its \textit{total mass} across much of cosmic time. Matter that was once labeled ``background" increasingly becomes labeled ``perturbation" and then ``halo" while the radial distribution of the domain's mass evolves, but those changes in labeling have only minor effects on the overall potential well, except near the origin, where virialization limits the central halo's matter density.

\section{Asymmetry and Centering}
\label{sec:Asymmetry}

Bound domains in the real universe are not spherical, of course. Their edges outline the ragged contours of large-scale structure, and precise tracking of those evolving edges requires numerical simulations. On one hand, the details of large-scale structure formation should not significantly alter the primary qualitative features of a spherical bound domain, such as its potential well depth, because those features depend much more strongly on a bound domain's total mass than on the distribution of matter within it. On the other hand, identifying the center of a bound domain before its central halo has significantly developed is a more subtle matter.

Early on, while the universe's matter distribution remains nearly uniform, a bound domain's boundaries are not obvious. The gravitational potential well around any given location during that time is nearly the same:
\begin{equation}
    \varphi (R) \approx \left[ \frac {\Omega_M} {2} (1 + z)^3 - \Omega_\Lambda \right]
           \frac {H_0^2 R^2} {2} + \varphi_0
    \; \; .
\end{equation}
This equation is nearly identical to equation (\ref{eq:phi_uniform}), except that the potential's zero point is set by the unspecified constant $\varphi_0$. The specific feedback energy input needed to push baryons to some distance $d$ from their original location (which is similar in magnitude to the specific kinetic energy of the Hubble flow at that distance) is therefore nearly independent of location. However, if we want to know whether a feedback event has been able to push baryons out of the bound domain in which the feedback event occurred, we need to know where the bound domain's edges are, so that we can compare the distance $d$ with the distance to an edge. 

In a cosmological numerical simulation, it is possible to determine a bound domain's edges by allowing the simulation to proceed into the far future. A set of collisionless particles that remains gravitationally bound in the far future defines a bound domain, and those particles can be traced backward in time. The boundaries of that particle distribution are then the early boundaries of the bound domain. In general, its boundaries will be asymmetric and also indistinct, for two reasons. First, two initially neighboring particles might not remain within the same bound domain, because gravitational interactions between the bound domain's particles can eject one subset while making the other subset more strongly bound. Second, neighboring structures outside of the bound domain exert tidal forces capable of removing particles that would otherwise remain bound \citep[e.g.,][]{Dunner_2006MNRAS.366..803D}. Particles close to the domain's edges are the most vulnerable to ejection, and so the probability of ejection increases with distance from the bound domain's center \citep[see, e.g.,][]{Behroozi_2013JCAP...06..019B}.

Given a bound domain's particle distribution, how should we go about identifying the domain's center? Here is one possibility: Add up the individual potential wells of all particles that end up in the \textit{asymptotic} bound domain. At a particular moment, the minimum of the resulting gravitational potential well can be considered the center of the bound domain. With that point designated $R=0$, adding the dark energy potential $\varphi_\Lambda (R)$ to the matter potential produces a bound domain potential with edges where $\varphi(R)$ reaches a maximum along each radial direction from the origin. Near those edges, the probability that baryons pushed out there might escape the bound domain can be set equal to the fraction of neighboring collisionless particles that do not remain in the asymptotic bound domain. 

Bear in mind, however, that asymptotic bound domains are well defined only in universe models with constant dark-energy density (i.e. with a cosmological equation-of-state parameter $w = -1$). If the dark-energy density increases with time ($ w < -1$), then bound domains gradually erode as dark energy's repulsion increases, and eventually vanish altogether. And if dark energy decays with time ($w > -1$), then there is not a one-to-one relationship between a bound domain's mass and its asymptotic radius.

\section{Open Questions}
\label{sec:Questions}

The preceding calculations were motivated by three open questions about galaxy groups and clusters: 
\begin{enumerate}

    \item Why do galaxy clusters have baryonic mass fractions similar to the cosmic mean while galaxy groups do not?

    \item How far do a group's baryons go when feedback pushes them beyond $R_{\rm 200c}$?

    \item What does the spatial distribution of baryons around groups tell us about the magnitude and timing of black-hole feedback?

\end{enumerate}
So far, this article's calculations fall somewhat short of answering those questions, but they do illuminate a way forward, that the next few sections will describe.

\section{Baryonic Binding Energy}
\label{sec:BaryonBinding}

At least one part of the answer to the first open question seems obvious: Baryons that end up in galaxy clusters are more strongly bound to their domains than the baryons end up in galaxy groups, even early in time, long before either clusters or groups have formed. Apparently, cumulative feedback energy output from supermassive black holes that will end up in galaxy groups is similar to the initial binding energy of baryons in a group-scale bound domain, but cumulative black-hole feedback output in a cluster-scale bound domain falls short of the baryonic binding energy.

To make the comparison between groups and clusters more quantitative, we would like to know the amount of feedback energy needed to eject \textit{just the baryons} from a bound domain of a given mass. The following subsection works out an estimate of that energy. Readers not interested in those details can skip ahead to the next subsection.

\subsection{Binding Energy Accounting}
\label{sec:Accounting}

We will start with the spherical shell model. Before ejection, the baryons in each shell $i$ have a specific kinetic energy
\begin{equation}
    \frac {\dot{R}_i} {2} =
                        \varepsilon_i  
                        + \frac {G M_i} {R_i} 
                        + \frac {H_0^2 \Omega_\Lambda} {2} R_i^2
                        \; \; .
\end{equation}
Now imagine removing all the baryons shell-by-shell, from the outside in. Lifting the baryons in shell $i$ away from the interior mass $M_i$ and out to $R_\infty$ requires a specific energy
\begin{equation}
    G M_i \left( \frac {1} {R_i} - \frac {1} {R_\infty} \right)
    \; \; .
\end{equation}
As the lifted baryons move from $R_i$ to $R_\infty$, work done by dark energy adds a specific energy
\begin{equation}
    \frac {H_0^2 \Omega_\Lambda} {2} \left( R_\infty^2 - R_i^2 \right)
    \; \; .
\end{equation}
Meanwhile, whatever is lifting the baryons to $R_\infty$ must do additional work against the gravitational attraction of the exterior mass shells between $R_i$ and $R_\infty$, amounting to a specific energy
\begin{equation}
    - \: ( 1 - f_{\rm b} ) \, \varphi_{\rm shells} (R_i,t)
    \; \; .
\end{equation}
The prefactor $(1 - f_{\rm b})$ here accounts for prior removal of baryons from the overlying shells. 

Combining these quantities gives the minimum amount of specific energy input needed to unbind the baryons in shell $i$:
\begin{equation}
   \Delta \varepsilon_i 
        \; = \: - \: \varepsilon_i 
             \: - \: \frac {G M_i} {R_\infty}
             \: - \: \frac {H_0^2 \Omega_\Lambda} {2} R_\infty^2
             \: - \: ( 1 - f_{\rm b} ) \, \varphi_{\rm shells} (R_i,t) 
             \; \; .
\end{equation}
The only time-dependent term on the right is the one accounting for the mass shells exterior to shell $i$. The others add up to the specific binding energy of shell $i$ to the mass interior to it.

Multiplying $\Delta \varepsilon_i$ by $f_{\rm b} (M_i - M_{i-1})$ and summing over all shells gives the total feedback energy needed to eject the bound domain's baryons. The $\varepsilon_i$ terms contribute $- f_{\rm b} E_{\rm bd}$ to the total, where
\begin{equation}
    E_{\rm bd} \equiv \sum_i  \left( M_i - M_{i-1} \right) \varepsilon_i
\end{equation}
is the bound domain's total initial energy. To represent the contributions of the terms including $R_\infty$, we can define
\begin{equation}
    E_\infty 
        \: \equiv \:  
            \frac {G M_\infty^2} {R_\infty}
        \: = \: 
            \sum_i \left( \frac {G M_i} {R_\infty} 
                            + \frac {H_0^2 \Omega_\Lambda} {2} R_\infty^2 \right)
                    \left( M_i - M_{i-1} \right)
            \; \; .
\end{equation}
To represent the contributions of exterior mass shells, we define
\begin{equation}
    E_\varphi (t) \: \equiv \: \sum_i \left( M_i - M_{i-1} \right)
                                \, \varphi_{\rm shells} (R_i,t) 
        \; \; .
\end{equation}
At time $t$, the total feedback energy needed to eject the bound domain's baryons is then
\begin{equation}
    E_{\rm b} (t) 
        = - f_{\rm b} \left[ E_{\rm bd} + E_\infty + ( 1- f_{\rm b} ) E_\varphi (t) \right]
    \; \; .
\end{equation}
This expression assumes all the baryons are still in their original mass shells at time $t$. However, feedback does not \textit{instantaneously} remove a bound domain's baryons, and so $E_{\rm b}$ should be considered just an estimate, not an exact feedback energy requirement.

Before proceeding, we need to address another complication: The potential well $\varphi_{\rm shells} (R_i,t)$ is undefined for shells that have already collapsed into the central halo by time $t$. To account for that complication, we can separate $E_\varphi$ into three parts:
\begin{equation}
    E_\varphi
        = E_{\varphi,{\rm shells}}
            + M_{\rm halo} \varphi_{\rm shells}(R_{\rm 200c})
            + E_{\varphi,{\rm halo}}
            \; \; .
\end{equation}
The first part is simply a sum over all the shells outside of the one (shell $j$) that has most recently collapsed:
\begin{equation}
    E_{\varphi,{\rm shells}} 
        \: \equiv \: \sum_{i > j} \left( M_i - M_{i-1} \right)
                                \, \varphi_{\rm shells} (R_i) 
        \; \; .
\end{equation}
The second part represents the energy required to lift the halo's mass, $M_{\rm halo}$, out of the collective potential well of the overlying shells, $\varphi_{\rm shells}(R_{\rm 200c})$. The third part represents the integral of $\rho_{\rm halo} \varphi_{\rm halo}$. For that third part, we will use an approximation based on integrating the density profile and potential well of the Hernquist halo model:
\begin{equation}
    E_{\varphi,{\rm halo}} 
        \: = \: \int 4 \pi R^2 \rho_{\rm halo} \varphi_{\rm halo} \, dR
        \; \approx \; \frac {G M_{\rm halo}^2} {R_\infty + R_H}
                    - \frac {G M_{\rm halo}^2} {3 R_H}
        \; \; .
\end{equation}
A more realistic approach would account for how $E_{\varphi,{\rm halo}}$ depends on the halo's mass accretion history.

\begin{figure}[!t]
    \centering
    \includegraphics[width=\linewidth]{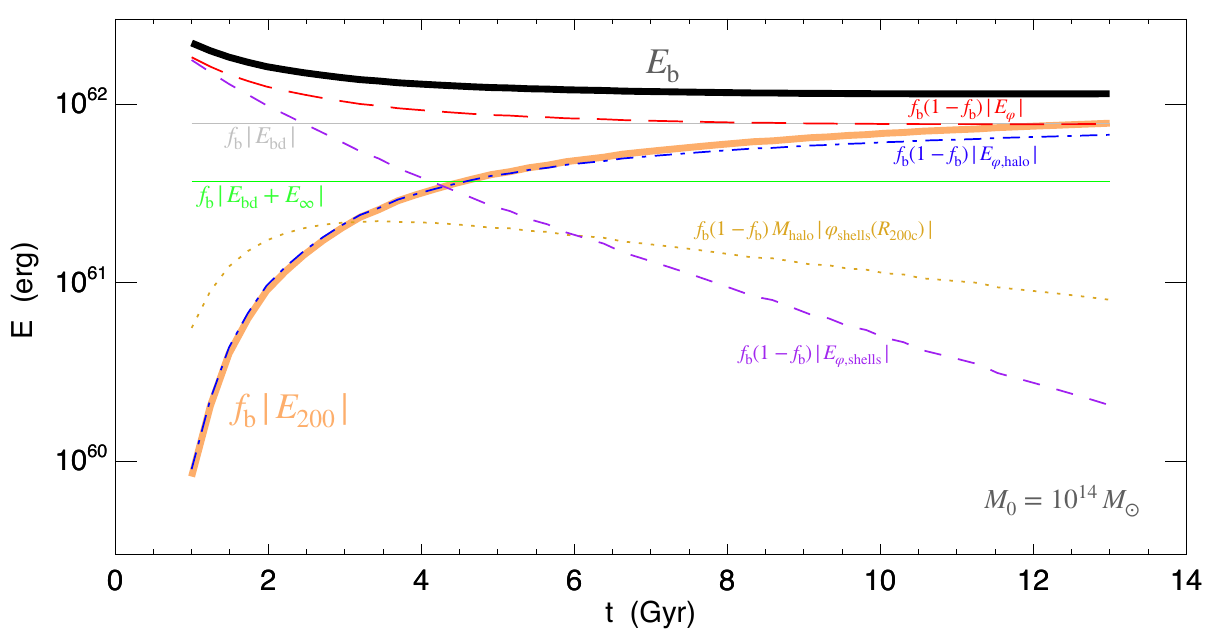}
    \caption{Evolution of baryonic binding energy in a group-scale bound domain with a central halo mass $M_0 = 10^{14} \, M_\odot$ at $z=0$. A thick black line shows the energy input $E_{\rm b}$ needed to unbind all of the bound domain's baryons at time $t$. A thick orange line shows the approximation $f_{\rm b} |E_{\rm 200}|$, representing the energy input needed to unbind just the central halo's baryons. The other lines show how various subcomponents of $E_{\rm b}$ evolve: The solid grey line is $f_{\rm b}$ times the magnitude of the bound domain's total energy. The solid green line shows $f_{\rm b}$ times the magnitude of the bound domain's binding energy. The long-dashed red line shows $f_{\rm b} (1 - f_{\rm b}) |E_\varphi|$, and the remaining lines show the contributions to $f_{\rm b} (1 - f_{\rm b}) |E_\varphi|$ coming from just the halo (dot-dashed blue line), lifting of halo baryons through the mass shells exterior to the halo (dotted gold line), and just the exterior shells (dashed purple line). }
    \label{fig:bd_binding_evol}
\end{figure}

Figure \ref{fig:bd_binding_evol} shows how these contributions to the total baryonic binding energy $E_{\rm b}$ evolve with time within the group-scale bound domain from Section \ref{sec:Example}.
Several colorful lines represent the terms that add up to $E_{\rm b}$. For reference, the group-scale bound domain has a total energy $E_{\rm bd} = -4.9 \times 10^{62} \, {\rm erg}$. 

%The thick black line near the top of Figure \ref{fig:bd_binding_evol} shows the energy input $E_{\rm b} (t)$ needed to unbind the bound domain's baryons while leaving the rest of the matter in place. A flat green line shows the energy $f_{\rm b} |E_{\rm bd} + E_\infty| = 3.7 \times 10^{61} \, {\rm erg}$ needed to unbind the baryons from the matter \textit{interior} to them. A dashed red line shows the energy $f_{\rm b} (1-f_{\rm b}) |E_\varphi|$ needed to unbind the baryons from the non-baryonic mass shells \textit{exterior} to them. That energy changes with time, and adding it to the green line gives the black line. 

%Three more lines in Figure \ref{fig:bd_binding_evol} represent the three components contributing to the dashed red line. A dot-dashed blue line shows the contribution $f_{\rm b} (1-f_{\rm b}) |E_\varphi|$ from just the halo. It is $\sim 10^{60} \, {\rm erg}$ at $t \sim 1 \, {\rm Gyr}$ and rises to nearly $10^{62} \, {\rm erg}$ by the present time. A dotted gold line shows the additional energy needed to lift the halo's baryons beyond the mass shells exterior to the halo. A dashed purple line shows the energy needed to lift baryons outside the halo beyond the mass shells exterior to their original shells.

\subsection{Binding Energy Early in Time}

Now back to the big picture: The most important thing to notice about Figure \ref{fig:bd_binding_evol} is the large difference between the energy input $E_{\rm b}$ needed to unbind \textit{all} of a bound domain's baryons (thick black line) and the energy input needed to unbind just the central halo's baryons (thick orange line). It can be more than two orders of magnitude early in time. Here, we are using $f_{\rm b} E_{\rm 200}$, in which $E_{\rm 200c} \equiv - G M_{\rm halo}^2 / 2 R_{\rm 200c}$, as our estimator for the halo's baryonic binding energy.  Clearly, using an estimator like that to assess whether early feedback episodes can unbind baryons from a group-scale environment is likely to lead to incorrect conclusions.

Finally, returning to the comparison between galaxy clusters and galaxy groups, notice that going from a group-scale mass of $10^{14} \, M_\odot$ to a cluster-scale mass of $10^{15} \, M_\odot$ increases a bound domain's baryonic binding energy by a factor of $10^{5/3} \approx 50$. Unbinding the baryons from a cluster-scale bound domain therefore requires $\sim 10^{64} \, {\rm erg}$ of feedback energy input, similar to the entire rest-mass energy of a $10^{10} \, M_\odot$ black hole. While there is no absolute upper limit on a black hole's mass that would prevent accretion of baryons from supplying $\sim 10^{64} \, {\rm erg}$ of feedback energy, astronomers have not yet found supermassive black holes with masses more than a few times greater than $10^{10} \, M_\odot$. Therefore, the differing baryon mass fractions observed in groups and clusters probably reflect differences in fueling of black-hole feedback that do not scale as steeply as $M_0^{5/3}$ all the way up to $10^{15} \, M_\odot$.

\section{Baryon Closure}
\label{sec:Closure}

Looking at cumulative feedback from a bound-domain perspective also helps to address the second open question, about how far a galaxy group's baryons go after feedback pushes them beyond $R_{\rm 200c}$. It is now being asked by both theorists and observers but is far from being definitively answered. On the theoretical side, black hole feedback in the current generation of cosmological simulations typically lifts baryons associated with halos of mass $\sim 10^{13}$--$10^{14} \, M_\odot$ out to $\sim 2$--4 times $R_{\rm 200c}$ \citep{Ayromlou_2023MNRAS.524.5391A}. In some simulations feedback pushes them to even greater altitudes \citep{Sorini_2022MNRAS.516..883S}. Observational constraints on baryon lifting are more challenging to obtain, but recent analyses of the kinetic Sunyaev-Zeldovich effect suggest the baryon distributions around galaxy groups are more extended than many simulations currently predict \citep[e.g.,][]{Amodeo_2021PhRvD.103f3514A,Hadzhiyska_2024arXiv240707152H}. 

\begin{figure}[!t]
    \centering
    \includegraphics[width=\linewidth]{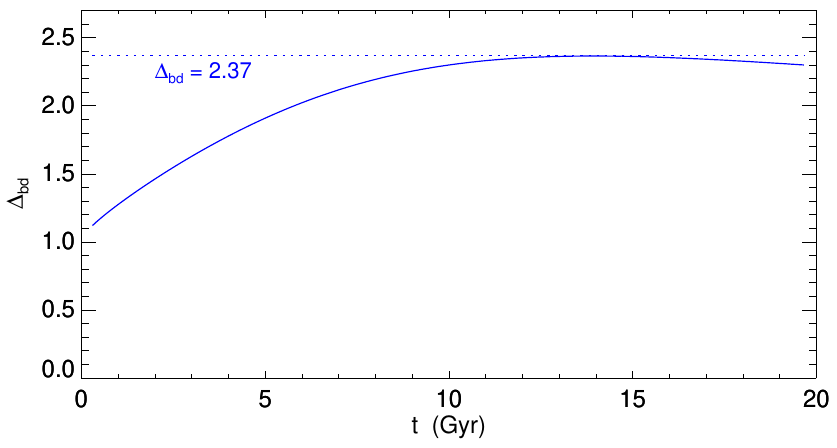}
    \caption{Evolution over time of a bound domain's density contrast factor $\Delta_{\rm bd}$ in a $\Lambda$CDM universe.}
    \label{fig:bd_delta}
\end{figure}

During the next several years, progress on the observational side may be able to tell us whether group-scale bound domains are closed systems of baryons, even though their baryonic mass fractions within $R_{\rm 200c}$ are $\lesssim 0.5 f_{\rm b}$. If a group's baryons do indeed remain within $\sim 4 R_{\rm 200c}$, as measured from the center of its dominant halo, then most of those baryons are likely to stay gravitationally bound to the group's bound domain, keeping the group's baryon cycle nearly closed. In Figure \ref{fig:bd_evol_labeled}, you can see that $R_{\rm bd}$ is between $4 R_{\rm 200c}$ and $6 R_{\rm 200c}$ across much of cosmic time. That is because a convenient rule of thumb applies as time progresses. In a $\Lambda$CDM universe, the critical density is $\rho_{\rm cr} = 3 H_0^2 [\Omega_{\rm M} (1+z)^3 + \Omega_\Lambda] / 8 \pi G$. A bound domain's mean matter density consequently exceeds $\rho_{\rm cr}$ by the factor 
\begin{equation}
    \Delta_{\rm bd} 
        \: \equiv \: \frac {3 M_{\rm bd}} {4 \pi R_{\rm bd}^3 \rho_{\rm cr}} 
        \: = \: \frac {2 \Omega_\Lambda} {\Omega_M (1+z)^3 + \Omega_\Lambda}
            \left( \frac {R_\infty} {R_{\rm bd}} \right)^3
        \; \; .
\end{equation}
As Figure \ref{fig:bd_delta} shows, that factor starts out near unity and peaks at $\Delta_{\rm bd} \approx 2.37$ near the present time \citep[see][for an analytical solution giving $R_{\rm bd}/R_\infty$ as a function of time]{Dunner_2006MNRAS.366..803D}. Thereafter, it gradually declines to $\Delta_{\rm bd} = 2$, implying that the ratio $R_{\rm bd} / R_{\rm 200c}$ asymptotically approaches $100^{1/3} \approx 4.6$. For the halo-growth model adopted in Figure \ref{fig:bd_evol_labeled}, the bound domain's current radius is around $4.8 R_{\rm 200c}$. More generally, we find
\begin{equation}
    \frac {R_{\rm bd}} {R_{\rm 200c}} 
        \: = \: \left( \frac {200 M_{\rm bd}} 
                        {\Delta_{\rm bd} M_{\rm halo}} \right)^{1/3}
        \approx \: 4.6 \left( \frac {M_{\rm bd}} 
                                {M_{\rm halo}} \right)^{1/3}
            \; \; .
\end{equation}
Baryons that stay within $\sim 4 R_{\rm 200c}$ are therefore likely to fall back toward a bound domain's central halo, if given enough time. However, whether or not the infalling gas reenters the central halo depends on the compressibility, and therefore on the specific entropy, of the gas that remains within $R_{\rm 200c}$.

\section{Quantifying Baryon Lifting}
\label{sec:Quantifying}

Further progress toward answering the second open question will inform answers to the third, concerning the magnitude and timing of black-hole feedback in regions of space that will become the universe's most massive halos. Here is where looking at things from a bound-domain perspective should be particularly helpful, for two reasons: (1) the excess energy responsible for reducing the baryon mass fraction of a single massive halo may have been released in numerous progenitor halos, long before they merged, and (2) cumulative feedback within a bound domain may impact  $\textit{all}$ of its progenitor halos early in time. 

For example, balancing the overall energy budget of a group-scale bound domain is far easier than tracking baryonic energy as it propagates through the hierarchical system of merging halos that will end up making a galaxy group. Initially, as black-hole feedback ejects baryons from smaller progenitor halos, much of that feedback energy may be in kinetic or thermal form. Over time, while gas motions convert that kinetic and thermal energy into gravitational potential energy, the total energy of the bound domain's baryons remains conserved. Then later, as those baryons settle toward the bound domain's most dominant halo, the excess energy that collective feedback has added limits how far they can descend. Treating the system as a single bound domain, rather than as many distinct halos, therefore directly links early feedback from many distinct halos to the gravitational potential energy of baryons within a single massive halo later in time.

When performing such calculations, one option for quantifying the excess gravitational energy of a bound domain's baryons is a measure I will call the \textit{baryon displacement energy:}
\begin{equation}
    E_{\rm D} 
     \: = \: \int_V \rho_b(\mathbf{r}) \varphi(\mathbf{r}) \, d \mathbf{r} 
        \: - \: f_{\rm b} \int_V \rho(\mathbf{r}) \varphi(\mathbf{r}) \, d \mathbf{r}
        \; \; .
\end{equation}
The first integral on the right is the gravitational potential energy of a bound domain's baryonic component, which has a density distribution $\rho_{\rm b} (\mathbf{r})$ within the bound domain's volume $V$. That volume doesn't need to be spherical. Ideally, it should be defined so that $\varphi$ is zero at the domain's border. But first, the simulated bound domain, its boundaries, and its potential well need to be identified and characterized as discussed in \S \ref{sec:Asymmetry}. The second term on the right then represents the energy input needed to lift baryons with an initial density distribution identical to the \textit{total} matter distribution out to the edge of the bound domain, where they are at least marginally unbound. Importantly, any baryons that have been pushed out of the bound domain, and do not contribute to the first integral, are implicitly treated as though they have zero potential energy.

This measure is defined so that $E_{\rm D}$ is initially negligible, before radiative cooling, galaxy formation, and feedback start to redistribute the bound domain's baryons. Once feedback significantly lifts those baryons, $E_{\rm D}$ becomes similar to the total feedback energy $E_{\rm fb}$ that has been added. And in extreme cases, complete removal of the baryons gives a value of $E_{\rm D}$ comparable to the initial binding energy of the bound domain's baryons ($E_{\rm b}$). However, there is not a simple one-to-one relationship between $E_{\rm D}$ and $E_{\rm b}$, because of how baryon redistribution alters the bound domain's \textit{collisionless} matter distribution, making its potential well shallower than it otherwise would have been. 

To go further, future studies will need to measure how both $E_{\rm D}$ and $E_{\rm fb}$ evolve within simulated bound domains, to see how closely they track each other. Calibrating that relationship will show how observational constraints on $E_{\rm D}$ provide information about $E_{\rm fb}$. Measuring how $E_{\rm D}$ scales with scales with halo mass in the low-redshift universe, can then inform us about black-hole feedback earlier in time. Such comparisons, particularly the scaling of $E_{\rm D}$ and $E_{\rm fb}$ with halo mass, may also reveal whether black-hole mass growth within the bound domains that will eventually become galaxy groups and clusters differs from black-hole mass growth in more isolated high-redshift halos of similar mass.

\section{Summary}

This tutorial article has aimed at sparking and supporting a broader conversation about \textit{bound domains}---the term it proposes for the gravitationally bound islands that will eventually become isolated from each other as the universe exponentially expands. It focuses on bound domains because they are fundamental to any conversation about whether feedback from either supernovae or black holes can unbind a halo's baryons. Early in time, while a halo's mass is small, it might be easy for feedback to expel baryons from the halo, but not from the halo's bound domain. And if that happens, then the expelled baryons will eventually reaccrete onto the bound domain's central halo. 

To support assessments of how and when that happens, sections \ref{sec:Dynamics}, \ref{sec:Well}, and \ref{sec:Evolution} set up the apparatus needed for calculating the evolution of an idealized cosmological potential well. Section \ref{sec:Example} then presents a group-scale example, showing that the overall depth of a bound domain's potential well does not evolve very much after the universe reaches an age of $\sim 1$~Gyr. Section \ref{sec:PertVsBack} explains this result, which might surprise some readers, in terms of the complementary roles of the background matter density and a growing perturbation. Section \ref{sec:Asymmetry} considers how bound domains in the real universe might differ from the idealized model.

The rest of the article discusses how viewing the universe in terms of bound domains may help to answer three open questions about galaxy groups and clusters, listed in section \ref{sec:Questions}. Section \ref{sec:BaryonBinding} estimates the baryonic binding energy of a spherical bound domain, showing that it can be more than two orders of magnitude larger than the baryonic binding energy of its central halo while the first supermassive black holes were growing. Section \ref{sec:Closure} considers how far baryons expelled from halos can go, while still remaining gravitationally bound. Section \ref{sec:Quantifying} proposes a metric that ties the spatial distribution of a bound domain's baryons to cumulative feedback energy output within the bound domain.

I hope these simple calculations will encourage others to develop a more sophisticated understanding of baryon distributions within bound domains using cosmological numerical simulations. 
\\

Partial support from this work came from the NSF through grant AAG-2106575.

\bibliography{BoundDomains}{}
\bibliographystyle{aasjournal}

\end{document}